\documentstyle[12pt]{article}
\renewcommand{\theequation}{\thesection.\arabic{equation}}
\begin{document}
\begin{titlepage}\vskip-1cm\hfill SLAC--PUB--6403

\hfill PITHA 93/43

\hfill December 1993

\hfill T/E

\hfill hep-ph/9312210
\begin{center}\ \ \
 \vskip 0.3cm
 {\LARGE {\bf Tracing CP violation in the \vskip 0.3cm
   production of top quark pairs by \vskip 0.3cm
multiple TeV proton--proton collisions
\footnote{Work supported by the Department of Energy, contract
 DE-AC03-76SF00515}}}
 \vskip 0.5cm
Werner Bernreuther \vskip 0.3cm Institut f. Theoretische Physik\\
Physikzentrum, RWTH Aachen\\
52056 Aachen, Germany \vskip 0.5cm Arnd Brandenburg
\footnote{Supported by Max Kade Foundation} \vskip 0.3cm Stanford Linear
Accelerator Center, P.O. Box 4349\\Stanford, California 94309,
USA\vskip 2cm
{\bf Abstract}
\end{center}

We investigate the possibilities of searching for non-standard
CP violation in $pp\to
t\bar{t}X$ at multiple TeV collision energies.
A general kinematic analysis of the
underlying partonic production processes $gg\to t\bar{t}$ and
 $q\bar{q}\to t\bar{t}$ in terms of their density matrices
is given. We evaluate the CP-violating parts of these matrices
in two-Higgs doublet extensions of the standard model (SM)
and give results
for CP asymmetries at the parton level. We show that these asymmetries
can be traced by measuring suitable observables constructed
from energies and momenta
of the decay products of $t$ and $\bar{t}$. We find CP-violating
effects to be
of the order of $10^{-3}$ and show that possible contaminations
induced by SM interactions are savely below the expected
signals.

\end{titlepage}

\section{Introduction}
\setcounter{footnote}{0}
A high energy and high luminosity proton-proton collider, such as the
planned Large Hadron Collider (LHC) at CERN, would be capable of
producing millions of top and antitop quarks. This would offer the
unique possiblity to explore in detail the physics of these quarks --
which have not been discovered yet, but for whose existence there is
indirect evidence \cite{Dallas}. Specifically, since  the top is known
to be heavy, $m_t > 113$ GeV \cite{CDF}, precision studies based on
large samples of $t\bar{t}$ events may serve as a probe,
through sizeable top-Yukawa couplings, to
the electroweak symmetry breaking sector (Higgs sector for short).
This sector may have a richer structure than the one
conceived in the standard model (SM) --- as is the case in many
of its extensions.
As a consequence a number of new phenomena may exist. A particularly
intriguing one is
a new ``source" of CP violation provided by the Higgs
                          sector\footnote{This source
was shown to be of interest in attempts to explain the
cosmological baryon asymmetry \cite{Nelson}.} which is unrelated
to the Kobayashi-Maskawa phase \cite{KM}. This is possible already in
the two-Higgs doublet extensions of the SM
\cite{Higgsmodel}--\cite{eigen}:
here neutral Higgs boson exchange leads to CP-violating
effects in fermionic amplitudes, and these effects would show up
most pronouncedly in reactions involving top quarks
\cite{Schr}--\cite{Bernbra}. The subject of this paper is to
investigate in detail\footnote{A short account of our work
was given in \cite{Bernbra}.}
                      the manifestation and the magnitude
       of neutral Higgs particle
CP violation in $pp\to t\bar{t}X$.
(For other studies on CP violation in top
quark production and decay see \cite{top5}--\cite{top4a}.)

The outline of our paper is as follows: In section 2 we give a general
kinematic analysis of the reactions $gg\to t\bar{t}$ and $q\bar{q}\to
t\bar{t}$ which are the leading partonic processes in $pp\to t\bar{t}X$.
We study these reactions in terms of their production density matrices
and describe the properties of these matrices under various symmetry
transformations including CP transformations. In section 3 we evaluate
the CP-violating parts of
the density matrices in a specific model, namely the two-Higgs doublet
extensions of the SM with CP-nonconserving neutral Higgs boson exchange.
Correlations which are sensitive to
                         CP violation at the parton level are
identified and results for their expectation values are presented. In
section 4 we show that these CP asymmetries can be traced in
$pp\to t\bar{t}X$ by looking at simple observables which involve
energies and/or momenta of the decay products of $t$ and $\bar{t}$.
Moreover, possible contaminations by CP-conserving interactions are
discussed and shown to be much smaller than the expected signals.
In an appendix we list the analytic results of our calculations.
\newpage

\section{Production density matrices for $gg\to t\bar{t}$
and $q\bar{q}\to t\bar{t}$}

\noindent  Because
 a heavy top has an extremely short lifetime ($\tau_t< 10^{-23}
 s$ if $m_t> 100$ GeV), the polarization of and spin--spin correlations
 between $t$ and $\bar{t}$ are not severely diluted by hadronization
 \cite{Kuehn}. These are ``good" observables in the sense that effects
 involving the spins of $t$ and $\bar{t}$ can be treated
 perturbatively.
 Therefore we will discuss the reaction $pp\to t\bar{t}X$,
respectively the underlying partonic processes in terms
of production density matrices.
We will consider only unpolarized $pp$ collisions.
At LHC energies $t\bar{t}$ pairs are produced mainly by gluon
gluon fusion. This reaction dominates over quark antiquark
annihilation into $t\bar{t}$.
We will first discuss the (unnormalized)
production density matrix for the
reaction
$ g(p_1)+g(p_2) \to t(k_1)+\bar{t}(k_2) $
in the gluon--gluon center of mass system. It is defined by

\begin{eqnarray}  R^{g}_{\alpha_1\alpha_2,\ \beta_1\beta_2}({\bf p},
{\bf k})
 =  N_g^{-1}\sum_{{\rm colors,gluon\ spin}} & \langle
t(k_1,\alpha_1), \bar t(k_2,\beta_1)  \vert {\cal T}\vert
 g(p_1),g(p_2) \rangle^* \nonumber \\
 & \langle t(k_1,\alpha_2), \bar t(k_2,\beta_2) \vert {\cal T}\vert
g(p_1),g(p_2) \rangle\label{ggtt} \end{eqnarray}

\noindent where $\alpha,\ \beta$ are spin indices, ${\bf p}={\bf p}_1,\
                                                   {\bf k}=
{\bf k}_1$ and $N_g= 256$. We sum here over the gluon spins and colors
                                                      since we
are
only interested in analysing the polarizations of the $t$ and $\bar{t}$
                                                      and
their
spin--spin correlations.
The matrix $R^{g}$ can be decomposed
in the spin spaces of $t$  and $\bar{t}$ as follows:

\begin{eqnarray} R^{g}= A^g1\!{\rm l}\otimes 1\!{\rm l}+B^{g+}_{i}\sigma^i
\otimes 1\!{\rm l}+B^{g-}_{i}1\!{\rm l}
\otimes\sigma^i+C^g_{ij}\sigma^i\otimes\sigma^j.\label{Rot}\end{eqnarray}

\noindent The first (second) factor in the tensor products of the $2\times 2$
unit matrix
$1\!{\rm l}$ and of
                 the Pauli matrices $\sigma^i$ refers to the $t$
                         $(\bar{t})$ spin
space.

Because of rotational invariance, the functions $B^{g\pm}_{i}$ and $C_{ij}$ can
be further decomposed:

\begin{eqnarray}B^{g\pm}_{i}=&b_{g1}^{\pm}\hat{p}_i+b_{g2}^{\pm}
\hat{k}_i+b_{g3}^{\pm}\hat{n}_i \nonumber \\
 C_{ij}=&c_{g0}\delta_{ij}
+ \epsilon_{ij\ell}(c_{g1}\hat{p}_{\ell}+
c_{g2}\hat{k}_{\ell}+c_{g3}\hat{n}_{\ell}) \nonumber \\ &
+c_{g4}\hat{p}_i\hat{p}_j+
c_{g5}\hat{k}_i\hat{k}_j+c_{g6}(\hat{p}_i\hat{k}_j+
\hat{p}_j\hat{k}_i)\nonumber \\ & +c_{g7}(\hat{p}_i\hat{n}_j+
\hat{p}_j\hat{n}_i)+c_{g8}(\hat{k}_i\hat{n}_j+
\hat{k}_j\hat{n}_i).\label{BC}\end{eqnarray}

\noindent Here the hat denotes a unit vector and ${\bf n}={\bf p}\times {\bf
k}$. The
structure functions $A^g$, $b_{gi}^{\pm}$ and $c_{gi}$ depend only on $\hat
s=(p_1+p_2)^2$
and on
    the cosine of the scattering angle, $z=\hat{\bf p} \cdot \hat {\bf k}$.

 Next we discuss the properties of $R^g$ under various
                  symmetry
transformations.
Since the initial $gg$ state is Bose symmetric, $R^g$ must satisfy

\begin{eqnarray} R^{g}(-{\bf p},{\bf k})=R^{g}({\bf p},{\bf k}).\end{eqnarray}

\noindent The initial $gg$ state, when averaged over colors and spins, is a CP
eigenstate in its
center of mass system. It is therefore possible to classify the
                       individual terms
in $R^g$
according to their CP transformation properties. If the interactions
were CP-invariant,
the matrix $R^g$ would have to satisfy

\begin{eqnarray} R^{g}_{\beta_1\beta_2,\ \alpha_1\alpha_2}({\bf p},{\bf k})\
=R^{g}_{\alpha_1\alpha_2,\ \beta_1\beta_2}({\bf p},{\bf k}).\end{eqnarray}

\noindent In table 1 we give a complete list of the transformation
                                    properties of the
structure functions under P, CP,
 and exchange of the initial gluons (``Bose''). It is also instructive
to collect the properties of these functions under time reversal (T)
and CPT transformations neglecting, just for this purpose,
                                   non-hermitean parts of the
scattering matrix. To give an example, table 1 is then to be read
as follows: for a T-invariant interaction one has
$b_{g3}^{\pm}(z) = -b_{g3}^{\pm}(z)$, i.e., $b_{g3}^{\pm} = 0$
only at the Born level, whereas at higher orders absorptive parts
render this function non-zero.
\par
\noindent Because of Bose symmetry, the structure functions $A^g,\
b_{g2}^{\pm},\
c_{g0},\ c_{g2},\ c_{g4},\ c_{g5}$ and $c_{g7}$ are even functions of $z$, the
other functions are odd in $z$.

The contributions to $R^g$ can be decomposed into a CP--even and a CP--odd
part:

\begin{eqnarray} R^{g}=R^{g}_{\rm
even}+R^{g}_{CP}.\label{EVENODD}\end{eqnarray}

\noindent  As can be read off from
  table 1, the CP--even term $R^{g}_{\rm even}$ in general
 has the following structure:

\begin{eqnarray} R^{g}_{\rm even}= & A^g1\!{\rm l}\otimes 1\!{\rm l}+
(b_{g1}^{\rm even}\hat{p}_i+b_{g2}^{\rm even}
\hat{k}_i+b_{g3}^{\rm even}\hat{n}_i)(\sigma^i
\otimes 1\!{\rm l}+1\!{\rm l}\otimes \sigma^i) \nonumber \\
& + (c_{g0}\delta_{ij}+c_{g4}\hat{p}_i\hat{p}_j+
c_{g5}\hat{k}_i\hat{k}_j+c_{g6}(\hat{p}_i\hat{k}_j+
\hat{p}_j\hat{k}_i)\nonumber \\ & +c_{g7}(\hat{p}_i\hat{n}_j+
\hat{p}_j\hat{n}_i)+c_{g8}(\hat{k}_i\hat{n}_j+
\hat{k}_j\hat{n}_i))\sigma^i\otimes\sigma^j .\label{EVEN}
\end{eqnarray}

\noindent Nonzero $b_{g1}^{\rm even},\ b_{g2}^{\rm even}$, $c_{g7},\ c_{g8}$
 can be induced only by
parity-violating interactions, $c_{g7},\ c_{g8}$ need in addition
                                                         absorptive
parts
in the scattering amplitude
when the interactions are CPT-invariant.
The structure function $b_{g3}^{\rm even}$ can only
get contributions from absorptive parts induced by parity-invariant
interactions.

\noindent The CP--odd term $R^{g}_{CP}$ reads

\begin{eqnarray} R^{g}_{CP}= &
(b_{g1}^{CP}\hat{p}_i+b_{g2}^{CP}
\hat{k}_i+b_{g3}^{CP}\hat{n}_i)(\sigma^i
\otimes 1\!{\rm l}-1\!{\rm l}\otimes \sigma^i) \nonumber \\
& + \epsilon_{ijk}(c_{g1}\hat{p}_i+
c_{g2}\hat{k}_i+c_{g3}\hat{n}_i)
\sigma^j\otimes\sigma^k .\label{ODD}\end{eqnarray}

\noindent CP-violating interactions which are also
               parity-violating can give contributions to
$b_{g1}^{CP},\ b_{g2}^{CP},\ c_{g1},\ c_{g2}$. Non\-zero $b_{g1}^{CP},\
b_{g2}^{CP}$
require in addition
        absorptive parts. C- and CP-violating interactions can induce
non--vanishing
structure functions $b_{g3}^{CP},\ c_{g3}$, where $c_{g3}\neq0$
requires in addition absorptive
parts.

The above discussion of the transformation properties of the structure
functions holds to all orders of pertubation theory.

The production density matrix $R^q$ for $q\bar{q}\to t\bar t$ is defined
 in complete analogy
to (\ref{ggtt}) as

\begin{eqnarray}  R^{q}_{\alpha_1\alpha_2,\ \beta_1\beta_2}({\bf p},
{\bf k})
 =  N_q^{-1}\sum_{{\rm colors},q\bar{q}\ {\rm spins}} &\langle
t(k_1,\alpha_1), \bar t(k_2,\beta_1)  \vert {\cal T}\vert
 q(p_1),q(p_2) \rangle^*  \nonumber \\
 &\langle t(k_1,\alpha_2), \bar t(k_2,\beta_2) \vert {\cal T}\vert
q(p_1),q(p_2) \rangle,\label{qqtt} \end{eqnarray}

\noindent where $N_q=36$. The decomposition of $R^q$ in the
spin spaces of $t$
and $\bar{t}$ is exactly the same as for $R^g$ (equ. (\ref{Rot})        ,
(\ref{BC})) as is the splitting into CP--even and odd terms (equ.
(\ref{EVENODD})-- (\ref{ODD})).
The transformation properties of the structure
functions $A^q,\ldots,c_{q8}$ of $R^q$ are the same as the respective ones for
$A^g,\ldots,
c_{g8}$ of $R^g$ given in table 1. Thus all conclusions derived from these
transformation
properties
--- except those from Bose symmetry, of course --- are also valid for the
structure functions
of $R^q$.
\section{CP violation and density matrices
in two-Higgs doublet models}
\setcounter{equation}{0}
Up to now our discussion has been independent of any specific model.
Suffice it to say that the Kobayashi-Maskawa mechanism of CP violation
\cite{KM} induces only tiny effects in the flavor-diagonal reactions
of sect. 2.
In the following we will concentrate on CP-violating effects generated
 by two--Higgs doublet extensions of the SM with CP violation
in the scalar potential \cite{Weinberg}. We briefly
recall the features of these models relevant for us. CP
violation  in the scalar potential
           induces mixing of CP--even and --odd scalars, thus leading
to three physical mass eigenstates $|\varphi_j\rangle\ (j=1,2,3)$
with no definite CP parity. That means, these bosons couple both to
scalar and pseudoscalar fermionic currents. For the top quark these
couplings are (in the notation of \cite{Schr}):

\begin{equation} {\cal
L}_{Y}=-(\sqrt{2}G_F)^{1/2}\sum_{j=1}^3(a_{jt}m_t\bar{t}t+
\tilde{a}_{jt}m_t\bar{t}i\gamma_5 t)\varphi_j, \end{equation}
\noindent where $G_F$ is Fermi's constant, $m_t$ is the top mass,
\begin{equation} a_{jt}=d_{2j}/\sin \beta,\ \ \
\tilde{a}_{jt}=-d_{3j}\cot \beta, \end{equation}

\noindent $\tan \beta=v_2/v_1$ is the ratio of vacuum expectation
values of the two doublets, and $d_{2j},\ d_{3j}$ are the matrix
elements of a $3\times 3$ orthogonal matrix which describes the mixing
of the neutral states \cite{Schr}.
                     In the following we assume that the couplings
and masses of $\varphi_{2,3}$ are such that their effect on all
quantities discussed below is negligible. Then the measure of CP
violation generated by  $\varphi\equiv\varphi_1$ exchange in
flavor--diagonal reactions like $gg\to t\bar{t},\ q\bar{q}\to t\bar{t}$
is
\begin{equation} \gamma_{CP}\equiv -a\tilde{a}=d_{21}d_{31}\cot\beta
/\sin \beta, \end{equation}
\noindent where we have put $ a=a_{1t},\ \tilde{a}=\tilde{a}_{1t}$.
So far, data from low energy phenomenology, in particular the
experimental upper bounds on the electric dipole moments of the
neutron \cite{neutron} and of the electron \cite{electron} do not
severely constrain this parameter: $\gamma_{CP}$ may be of order one.
We note here that the couplings of the $\varphi_j$ to quarks and
leptons induce CP violation already at the Born level. The especially
interesting case of a  Higgs boson
                       $\varphi$ decaying into $t\bar{t}$ was shown in
\cite{Bernbra} to lead to CP-violating spin--spin correlations
            which may be as large as 0.5.
(For other discussions of the CP properties of neutral Higgs
bosons see
\cite{Ma}--\cite{endhiggs}.)

We will now discuss the structure of the matrices $R^{g}_{CP}$ and
$R^{q}_{CP}$
in these models.
               The Higgs boson contributions to the processes $gg\to
t\bar{t}$ and $q\bar{q}\to t\bar{t}$ discussed in section 2
are shown, together with the leading SM diagrams, in figs. 1, 2. Since
the CP-nonconserving neutral Higgs exchange is, in particular,
parity-violating, the relations
\begin{equation} R_{CP}^{g(q)}(-{\bf p},-{\bf k})=-R_{CP}^{g(q)}({\bf
p},{\bf k}) \end{equation}

\noindent hold as long as $R_{CP}$ results from interference
of these Higgs exchange amplitudes with amplitudes from
parity-invariant interactions.
This forces $b_{g3}^{CP}$, $c_{g3}$ and $b_{q3}^{CP}$,
$c_{q3}$ to be zero in these models. Furthermore, the virtual
intermediate gluon produced by annihilation of unpolarized $q$ and
$\bar{q}$ cannot have a vector polarization. Thus the contributions of
fig. 2 to $R^q$ are invariant with respect to the substitution ${\bf
p}={\bf p}_1\to -{\bf p}$. Hence the structures of $R^g_{CP}$ and
$R^q_{CP}$ are the same; the functions $b_{g1}^{CP}$, $c_{g1}$,
$b_{q1}^{CP}$, $c_{q1}$ of eqn. (\ref{ODD}) are odd
 under $z\to -z$, whereas
$b_{g2}^{CP}$, $c_{g2}$,
$b_{q2}^{CP}$, $c_{q2}$ are even functions of $z$.\\
                                                   The explicit results
for the matrices $R^g$ and $R^q$ evaluated from the diagrams of fig. 1,
2, respectively, are given in the appendix. The width
of $\varphi$ must be taken into account in the calculation of $R^g$ if
$\varphi>2m_t$, since in that case the contribution from fig. 1h
can become resonant. Because for $\varphi>2m_t$ the width of $\varphi$
is not very small as compared to its mass it is important to note that
the narrow width approximation cannot be applied in this case.\\
In view of the above discussion
it is now very easy to identify the correlations at the parton level
                                           which
trace the various CP--odd parts of the production density matrices. The
expectation value of an observable ${\cal O}$ for the respective parton
reactions is defined as

\begin{equation} \langle {\cal O}\rangle_i=
\frac{\int_{-1}^{1}dz{\mbox tr}(R^i{\cal O})}{4\int_{-1}^1dzA^i} \ \ \ \ \ \
(i=g,q) \end{equation}

\noindent Contributions of the functions $b_{g1,g2}^{CP}$, $b_{q1,q2}^{CP}$
are picked up by taking expectation values of

\begin{equation} \hat{\bf k}\cdot ({\bf s}_+-{\bf
s}_-)f_e(z),\label{a1} \end{equation}
\noindent or
\begin{equation} \hat{\bf p}\cdot ({\bf s}_+-{\bf
s}_-)f_o(z),\label{a2} \end{equation}
\noindent or linear combinations thereof, where ${\bf s}_+$, ${\bf
s}_-$ are the spin operators of $t$ and $\bar{t}$, respectively,
$f_e(z)$ is an even function of $z$ and $f_o(z)$ is odd in $z$. One has
for example

\begin{equation} \langle\hat{\bf k}\cdot ({\bf s}_+-{\bf
s}_-) \rangle_g=
\frac{4\int_{-1}^{1}dz(zb_{g1}^{CP}+b_{g2}^{CP})
}{4\int_{-1}^1dzA^g}, \end{equation}

\noindent and likewise for $\langle \hat{\bf k}\cdot ({\bf s}_+-{\bf
s}_-)\rangle_q$.

The result for the basic longitudinal polarization asymmetry
$ \langle\hat{\bf k}\cdot ({\bf s}_+-{\bf
s}_-) \rangle_g$ is plotted in fig. 3 as a function of the parton CM
energy for $m_t=150$  GeV and two values of the Higgs boson mass: The
dashed curve corresponds to $m_{\varphi}=100$ GeV and $\gamma_{CP}=1$.
For $m_{\varphi}$ of the order of $2m_t$ or larger, the shape of the
resulting graph depends, for fixed $m_{\varphi}$, on the strength of
the Higgs couplings $a,\ \tilde{a}$ and on the couplings to $W^+W^-,\
ZZ$ determining the width of the $\varphi$. (See eqn. (\ref{Gamma}) for
details.) For the solid line we have chosen $m_{\varphi}=350$ GeV,
$\vert a\vert=\vert\tilde{a}\vert=\gamma_{CP}=1$ and
$\Gamma_{\varphi}=47$ GeV. The asymmetry $\langle \hat{\bf k}\cdot
({\bf s}_--{\bf
s}_+)\rangle$ corresponds to the asymmetry $\Delta
N_{LR}=[N(t_L\bar{t}_L)-N(t_R\bar{t}_R)]/({\mbox{all}}\  t\bar{t})$
studied in \cite{Peskin}. We reproduce the numerical results of
\cite{Peskin} for $\Delta N_{LR}$ if we neglect $\Gamma_{\varphi}$ and
use $\gamma_{CP}=1/\sqrt{2}$  which corresponds to the parameter
Im$(A^2)=\sqrt{2}$ used in \cite{Peskin}.

The functions $c_{g1,g2},\ c_{q1,q2}$ generate nonzero expectation
values of the triple product correlations
\begin{equation}
 \hat{\bf k}\cdot ({\bf s}_+\times{\bf
s}_-) h_e(z),\label{ss1} \end{equation}

\begin{equation}
 \hat{\bf p}\cdot ({\bf s}_+\times{\bf
s}_-) h_o(z),\label{ss2} \end{equation}

\noindent and of their linear combinations, where $h_e$ and $h_o$ are
even and odd functions of $z$, respectively. For example,

\begin{equation} \langle\hat{\bf k}\cdot ({\bf s}_+\times{\bf
s}_-) \rangle_g=
\frac{2\int_{-1}^{1}dz(zc_{g1}+c_{g2})
}{4\int_{-1}^1dzA^g}. \end{equation}

\noindent In fig. 4 we plot this basic CP--odd and T--odd spin--spin
correlation with the same choice of parameters as in fig. 3. It reaches
values of up to about two percent.

For completeness, we show in figs. 5 and 6 the expectation values
$\langle\hat{\bf k}\cdot ({\bf s}_+-{\bf
s}_-) \rangle_q$ and $\langle\hat{\bf k}\cdot ({\bf s}_+\times{\bf
s}_-) \rangle_q$, respectively, again for the same choice of parameters
as in fig. 3. Here the CP asymmetries get smaller with growing Higgs
masses.

\section{CP observables for $pp\to
t\bar{t}X$ }
\setcounter{equation}{0}
The CP-violating spin-momentum correlations for $t$ and $\bar{t}$
of the previous section must be traced in the final states
                        into which
$t$ and $\bar{t}$ decay.
In this
section we discuss a few observables which allow to do this.
 The charged lepton from $t\to Wb\to \ell^+\nu_{\ell}b$ is
  an efficient analyzer of the top spin \cite{Jez}. We will therefore
  conider only decay chains where at least one of the
  top quarks decays semileptonically. We shall use the SM decay
density matrices as given in \cite{top9,top11}.

Observables in $pp\to t\bar{t}X$ cannot be classified as being even or
odd with respect to CP, because the initial state is not a CP
eigenstate. However, they can be classified as being T--even or T--odd
(i.e. even or odd under reflection of momenta and spins).
Their expectation values
      will in general be contaminated by contributions
from CP-conserving
interactions.

The asymmetries
            in the $t$ and $\bar{t}$ polarizations in the production
plane, as given in eqns. (\ref{a1}) and (\ref{a2}), translate into
T--even observables formed by energies and/or
                                            momenta of the final states.
As an example, we have investigated the expectation values of the
following two observables (another one was given in \cite{Peskin}):
\begin{equation} A_1=E_+-E_- \end{equation}
\begin{equation} A_2={\bf k}_{\bar{t}}\cdot {\mbox{\boldmath $\ell$}}^+-
{\bf k}_{t}\cdot {\mbox{\boldmath $\ell$}}^-.\end{equation}
\noindent Here $E_{\pm}$, ${\mbox{\boldmath $\ell$}}^{\pm}$ are the energies
and
momenta of the leptons in $t\to\ell^+\nu_{\ell}b$ and
$\bar{t}\to\ell^-\bar{\nu}_{\ell}\bar{b}$ in the laboratory frame and
${\bf k}_{t(\bar{t})}$ is the top (antitop) momentum in this system.

To measure $A_2$, one has to select events where the $t$ decays
leptonically and the $\bar{t}$ hadronically, which in principle
                                     allow to reconstruct the
$\bar{t}$ momentum \cite{Ladin},  and vice versa.
  We will give explicit expressions
for the expectation values of $A_1$ and $A_2$ below
                                              when we discuss
contaminations by CP-conserving interactions. In calculating these
expectation values we have used the
          narrow width approximation for the top: because
the top width is much smaller than its mass (in view of the
experimental upper bound on $m_t$ which is of the order of $200$ GeV),
the approximation of the on--shell production of $t$ and $\bar{t}$
followed by their weak decays yields a good description of the
reactions considered here. We also neglected CP violation in the decays
of $t$ and $\bar{t}$ (for comments on this, see \cite{Bernbra}). For
the parton distributions entering the calculation of expectation
values in $pp$ collisions we have used the parametrization of
\cite{DO}. In order to assess the statistical sensitivity of the
observables $A_1$ and $A_2$, we have computed the signal--to--noise
ratios $\langle A_{i}\rangle /\Delta A_{i}\ (i=1,2)$ for Higgs masses
 $100 \mbox{ GeV}\le m_{\varphi} \le 450 \mbox{ GeV}$ both for
$\sqrt{s}=15 \mbox{ TeV}$ (LHC) and $\sqrt{s}=40  \mbox{ TeV}$.
 Here $\Delta A_i=\sqrt{\langle A_i^2\rangle -\langle
A_i\rangle ^2}$ denotes the width of the distribution of $A_i$. We
present our results in figs. 7 and 8 for the same parameter set as used
in calculating the partonic asymmetries, that is, $a=-\tilde{a}=1,
\hat{g}_{VV}=1$ (for the definition of $\hat{g}_{VV}$ see appendix,
eqn. (\ref{gvv})). We integrate here over the whole phase space. Both
observables have signal--to--noise ratios of order $10^{-3}$. LHC
offers larger effects due to the fact that $\Delta A_{1,2}$ is larger
for $\sqrt{s}$= 40 TeV.

If both $t$ and $\bar{t}$ decay leptonically, one can look at the T--odd
observable
\begin{equation} T_2=({\bf b}-{\bf \bar{b}})\cdot ({\mbox{\boldmath
$\ell$}}^+\times
{\mbox{\boldmath $\ell$}}^-) \end{equation}

\noindent where ${\bf b},\ {\bf \bar{b}}$ denote the momenta of the $b$ and
$\bar{b}$ jets in the laboratory frame. (This observable was also
discussed in \cite{Bernbra}.) The expectation value of $T_2$ traces the
spin--spin correlations of (\ref{ss1}) and (\ref{ss2}). In fig. 9 we
show the signal--to--noise ratio as a function of the Higgs mass
for this observable, again with the same choice of parameters as
for $A_{1,2}$. The effect is also of the order of $10^{-3}$.

We will now discuss in some detail possible contaminations of the
observables $A_{1,2}$
                and $T_{2}$ due to CP-conserving interactions. Such
contaminations arise in particular
                     because the $pp$ initial state is not a CP
eigenstate. One can give general arguments why these contaminations
should be small. Most importantly, the dominant subprocess
is gluon fusion which does not induce any CP-conserving
contributions to our observables (cf. \cite{Bernbra} and below).
                                 Furthermore, T--odd observables like
$T_{2}$ do not receive contributions from CP-invariant interactions
at the Born level but only from absorptive parts. The main background
in this case comes from order $\alpha_s^3$ and order
$\alpha_s^2\alpha_{\rm weak}$ absorptive parts in $q\bar{q}\to
t\bar{t}$ which generate nonzero functions $b_3^{\rm even}$ in $R^q$.
However, numerical simulations show that these contributions are
smaller than $10^{-6}$, i.e. about three orders of magnitude smaller
than the signal shown in fig. 9. Potentially more dangerous are
CP--even contributions to $A_1$ and $A_2$, because, as will be shown
below, they can already be generated by weak interactions at the Born
level.

Integrating over the whole phase space we can
actually give explicit analytic formulae for the expectation values
of $A_1$ and $A_2$ in terms of the structure functions. This is very
illuminating for identifying possible contaminations. We
have carried out our calculations within
  the naive parton model (which neglects intrinsic transverse
momenta of the incoming partons) and restricted ourselves again to
$gg$ and $q\bar{q}$ initial states. Furthermore,
 we have used the narrow width approximation desribed above
 and have taken into account only the  SM  decays of $t$ and
$\bar{t}$. Then we find:

\begin{eqnarray} \langle A_1\rangle&=&\frac{1}{\sigma}\frac{1}{2s}
\frac{g(m_W^2/m_t^2)}{4\pi}\nonumber \\ & &
\Big\{\int_0^1dx_1\frac{N_g^p(x_1)}{x_1}
\int_0^1dx_2\frac{N_g^p(x_2)}{x_2}\int_{-1}^{1}dz\beta\frac{x_1+x_2}
{2\sqrt{x_1x_2}}\frac{E_1}{3}(zb_{g1}^{CP}+b_{g2}^{CP})\nonumber \\
& & + 2\int_0^1dx_1\frac{N_q^p(x_1)}{x_1}
\int_0^1dx_2\frac{N_{\bar q}^p(x_2)}{x_2}\int_{-1}^{1}dz\beta\Big [
\frac{x_1+x_2}{2\sqrt{x_1x_2}}\frac{E_1}{3}
(zb_{q1}^{CP}+b_{q2}^{CP})\nonumber \\& &+\frac{x_1-x_2}{2\sqrt{x_1x_2}}
\Big (E_1\beta zA^q+\frac{1}{3}([(1-z^2)m_t+z^2E_1]b_{q1}^{\rm even}
+E_1zb_{q2}^{\rm even})
\Big )\Big ]
\Big \},\label{A1} \end{eqnarray}
\vfill
\newpage

\begin{eqnarray} \langle A_2\rangle&=&\frac{1}{\sigma}\frac{1}{2s}
\frac{g(m_W^2/m_t^2)}{4\pi}\nonumber \\ & &
\Big\{\int_0^1dx_1\frac{N_g^p(x_1)}{x_1}
\int_0^1dx_2\frac{N_g^p(x_2)}{x_2}\int_{-1}^{1}dz\beta\Big [
-\frac{E_1^2\beta}{3}(zb_{g1}^{CP}+b_{g2}^{CP})\nonumber \\ & &
+\frac{(x_1-x_2)^2}{4x_1x_2}\frac{E_1\beta}{3}(1-z^2)((E_1-m_t)
zb_{g1}^{CP}+E_1b_{g2}^{CP})\Big ]
\nonumber \\
& & + 2\int_0^1dx_1\frac{N_q^p(x_1)}{x_1}
\int_0^1dx_2\frac{N_{\bar q}^p(x_2)}{x_2}\int_{-1}^{1}dz\beta\Big [
-\frac{E_1^2\beta}{3}(zb_{q1}^{CP}+b_{q2}^{CP})\nonumber \\ & &
+\frac{(x_1-x_2)^2}{4x_1x_2}\frac{E_1\beta}{3}(1-z^2)[(E_1-m_t)
zb_{q1}^{CP}+E_1b_{q2}^{CP}]\nonumber \\ & &+\frac{x_1^2-x_2^2}{4x_1x_2}
\frac{m_t}{3}([(1-z^2)E_1+z^2m_t]b_{q1}^{\rm
even}+m_tzb_{q2}^{\rm even})\Big ]
\Big \}.\label{A2} \end{eqnarray}

\noindent Here $\sigma$ is the total cross section for $pp\to t\bar{t} X
$, $s$ is the $ p p$
  collision energy squared, $g(y)=(1+2y+3y^2)/(2+4y)$,
$N_{g}^p$, $N_{q(\bar{q})}^p$  denote the gluon and
quark (antiquark)
distribution functions of the proton, $E_1$ is the energy of the
top quark in the partonic CM and $\beta=(1-m_t^2/E_1^2)^{1/2}$.

 Equations (\ref{A1}) and (\ref{A2}) exhibit several interesting
features:

\noindent - One can see explicitly that gluon fusion generates no
CP-even contributions to the observables.

\noindent - Quark--antiquark annihilation produces several
contaminations: In $\langle A_1\rangle$ a term $\sim
zA^q(z)$ appears which, after integrating over $z$,
         is nonzero only if $A^q(z)$
has a part which is odd in $z$; that is, if $q\bar{q}\to t\bar{t}$ has a
forward--backward asymmetry. Such an asymmetry is induced in order
$\alpha_s^3$. (In \cite{Peskin} this potential source of
 contaminations was discussed.)
Possibly more important are the terms $b_{q1}^{{\rm even}}$ and
$b_{q2}^{{\rm even}}$ which appear in both expectation values above
 because these terms can
be generated at the Born level via $q\bar{q}\to Z \to t\bar{t}$.
We calculated their contributions and found that for both observables
they are suppressed by more than two orders of magnitude in comparison
to the signals shown in figs. 7 and 8.

A future multiple TeV and high luminosity collider like the LHC
has the potential of producing more than $10^7$ $t\bar{t}$ pairs.
If it were for statistics alone detection of
                                effects of a few permil which we
found might be feasible. More detailed (Monte Carlo) studies
including judicious choices of phase space cuts are required in order
to explore the possiblity of enhancing the signals by some factor.
A crucial issue will eventually be whether detector effects can be
kept at the level of $10^{-3}$. \vfill \newpage

\section{Conclusions}
In this article we have studied the possibility of detecting
CP violation in top quark pair production at future hadron
colliders. We have given a general kinematic analysis of the underlying
dominant partonic
         subprocesses and identified the relevant CP asymmetries at the
parton level. We have further computed these asymmetries in
two-Higgs
doublet extensions of the SM where CP violation is generated
through neutral Higgs boson exchange. Whereas at the parton level these
models can induce asymmetries of the order of a few percent, realistic
observables built up from energies and/or
                                       momenta of the final states into
which the top quarks decay give signal--to--noise ratios of up to a few
$\times 10^{-3}$. Contaminations  by CP-conserving
interactions were shown to be much smaller than the signals. Since the
issue of CP violation is of fundamental interest detailed
investigations of the experimental feasibility of an observation of
these effects would certainly be worthwhile.
\vskip 1.0cm
\noindent {\bf Acknowledgments}
\vskip 0.5cm
\noindent A. B. would like to thank the SLAC
theory group for the hospitality extended to him.
\newpage

\noindent {\Large{\bf Appendix}}
\vskip 0.5cm
\setcounter{equation}{0}
\renewcommand{\theequation}{A.\arabic{equation}}
\noindent In this appendix we list our analytic results for the structure
functions of the $t\bar{t}$ spin density matrix $R^g$
defined in equ.
(\ref{ggtt})--(\ref{BC}) and decomposed into
                                           CP--even and CP--odd parts in
equ. (\ref{EVENODD})--(\ref{ODD}) and also the corresponding functions
in $R^q$ defined
in (\ref{qqtt}). All calculations are carried out in the two-Higgs
                                                            doublet
extensions of the SM described in section 3. The relevant Feynman
diagrams are shown in figures 1a--h for the process $gg\to t\bar{t}$
and in figures 2a,b for $q\bar{q}\to t\bar{t}$.\par $R^q$ is obtained
very easily: The CP--even part is determined to good approximation by
the Born diagram fig. 2a, whereas $R^q_{CP}$ results from the
interference of fig. 2b (with couplings $a\tilde{a}=-\gamma_{CP}$)
 with fig. 2a. The nonzero CP--even structure functions of $R^q$ read:

\begin{eqnarray} A^q& =& \frac{g_s^4 (\beta^2(z^2-1)+2)}{18}\nonumber \\
 c_{q0}&=& \frac{g_s^4 \beta^2(z^2-1)}{18}\nonumber \\
c_{q4}&=&\frac{g_s^4}{9}\nonumber \\
c_{q5}&=& \frac{g_s^4\beta^2}{9}\Big (\frac{\beta^2 z^2 E_1^2}{(E_1+m_t)^2}+1
\Big )\nonumber \\
c_{q6}&=& \frac{-g_s^4\beta^2 z E_1}{9( E_1+m_t)}. \end{eqnarray}

\noindent Here and in the following $g_s$ denotes the strong coupling
constant, $E_1$ is the energy of the top in the CM system of the
incoming partons and  $\beta=\sqrt{1-m_t^2/E_1^2}$. Recall that
$z=\hat{\bf p}\cdot \hat{\bf k}$.

\noindent The CP--odd contributions are:

\begin{eqnarray} b_{q1}^{CP}&=&\frac{-m_t^3\sqrt{2}G_F
\gamma_{CP} }{ 8\pi^2}
\frac{4 g_s^4 E_1\beta z}{9} {\rm Im}G(\hat s),\\
b_{q2}^{CP}&=&\frac{m_t^3\sqrt{2}G_F
\gamma_{CP} }{ 8\pi^2}
\frac{4 g_s^4 E_1\beta}{9} \Big [\frac{(z^2-1)\beta^2 E_1}{ E_1+m_t}+1\Big ]
{\rm Im}G(\hat s),\\
c_{q1}&=&\frac{m_t^3\sqrt{2}G_F
\gamma_{CP}}{8\pi^2}
\frac{4 g_s^4 E_1\beta z}{9} {\rm Re}G(\hat s),\\
c_{q2}&=&\frac{-m_t^3\sqrt{2}G_F
\gamma_{CP}}{8\pi^2}
\frac{4 g_s^4 E_1\beta}{9} \Big [\frac{(z^2-1)\beta^2 E_1}{E_1+m_t}+1\Big ]
{\rm Re}G(\hat s).\end{eqnarray}

\noindent Here $G_F$ is Fermi's constant and

\begin{eqnarray}G(\hat s)={-(m_{\varphi}^2 C_0(\hat
s,m_{\varphi}^2,m_t^2,m_t^2)
+B_0(\hat s,m_t^2,m_t^2)-B_0(m_t^2,m_{\varphi}^2,m_t^2))\over \hat{s}
\beta^2},\end{eqnarray}
\newpage
\noindent where
\begin{eqnarray}& & C_0(\hat{s},m_{\varphi}^2,m_t^2,m_t^2)=\nonumber \\
& &\int \frac{d^4 l}{i\pi^2}
\frac{1}{l^2-m_{\varphi}^2+i\epsilon}\frac{1}{(l+k_1)^2-m_t^2+i\epsilon}
\frac{1}{(l+k_1-p_1-p_2)^2-m_t^2+i\epsilon}.\label{C01}\end{eqnarray}

\noindent is a standard three-point scalar integral which can be
reduced to dilogarithms \cite{Hooft}.
          We note here that for the models of section 3 all structure
                                                            functions
are ultraviolet finite. In particular, the scalar two-point functions
$B_0$ show up in all our results only as differences of the form

\begin{eqnarray}& &B_0(q_1^2,m_1^2,m_t^2)-B_0(q_2^2,m_2^2,m_t^2)
= \nonumber \\& & -\int_0^1dx{\rm log}\Big [
\frac{x^2q_1^2+x(m_t^2-m_1^2-q_1^2)+m_1^2-i\epsilon}{
x^2q_2^2+x(m_t^2-m_2^2-q_2^2)+m_2^2-i\epsilon }\Big ].\end{eqnarray}

\noindent This completes our results for the matrix $R^q$.\par
The computation of $R^g$ is more involved since the contribution of
fig. 1h becomes resonant if $m_{\varphi}>2m_t$. (Fig. 1h actually
represents four amplitudes: two CP--conserving ones with couplings
$a^2$ and $\tilde{a}^2$, respectively, and two CP--violating ones with
couplings $a\tilde{a}$.) The width of $\varphi$ must therefore be taken
into account in the $\varphi$ propagator. We compute $\Gamma_{\varphi}$
by summing the partial widths for $\varphi\to W^+W^-,\ ZZ,\ t\bar{t}$ in
the two--Higgs doublet model which contains (3.1). At the Born level
only the $CP=+1$ component of $\varphi$ couples to $W^+W^-$ and $ZZ$.
The couplings are given by the respective SM couplings times the
factor

\begin{equation} \hat{g}_{VV}=(d_{11}\cos\beta+d_{21}\sin
\beta).\label{gvv}
\end{equation}
\noindent  Explicitly,

\begin{eqnarray} \Gamma_{\varphi}&=&\Gamma_W+\Gamma_Z+\Gamma_t \nonumber \\
\Gamma_W&=&\Theta(m_{\varphi}-2m_W)\frac{\hat{g}_{VV}^2m_{\varphi}^3\sqrt{2}G_F
\beta_W}{ 16\pi}\Big [\beta_W^2+12\frac{m_W^4}{
m_{\varphi}^4}\Big ]\nonumber \\
\Gamma_Z&=&\Theta(m_{\varphi}-2m_Z)\frac{\hat{g}_{VV}^2m_{\varphi}^3\sqrt{2}G_F
\beta_Z}{ 8\pi}\Big [\beta_Z^2+12\frac{m_Z^4}{
m_{\varphi}^4}\Big ]\nonumber \\
\Gamma_t&=&\Theta(m_{\varphi}-2m_t)\frac{3m_{\varphi}m_t^2\sqrt{2}G_F
\beta_t}{8\pi}(\beta_t^2a^2+\tilde{a}^2)\label{Gamma}
.\end{eqnarray}

\noindent Here we have used the notation
$\beta_{W,Z,t}=(1-4m_{W,Z,t}^2/m_{\varphi}^2)^{1/2}$.
In order to incorporate the resonance region we have
determined $R^g_{{\rm even}}$ from the squared Born amplitudes figs.
1a, 1b, the interference of fig. 1a with the CP--even amplitudes of fig.
1h, and the squared amplitudes of fig. 1h. We denote the Born
contributions by a lower index ``Born'' and the other two contributions
by a lower index ``resonance'' in the following. The results for the
nonzero structure functions of $R^g_{{\rm even}}$ are:
\vfill\eject

 \begin{eqnarray} A^g&=&A^g_{\rm Born}+A^g_{\rm resonance}\nonumber \\
A^g_{\rm Born}&=&
\frac{g_s^4 \left( 7 + 9\beta^2 z^2 \right)
 }{192 E_1^4
      \left( -1 +  \beta^2 z^2  \right) ^2}
      (E_1^4  +
       2E_1^2   m_t^2  -
       2 m_t^4 -
       2 \beta^2 E_1^2 m_t^2
         z^2  -  \beta^4 E_1^4   z^4 ) \nonumber \\
  A^g_{\rm resonance}&=& \frac{g_s^4}{
(\hat s-m_{\varphi}^2)^2+
\Gamma_{\varphi}^2m_{\varphi}^2} \Big \{-\frac{1}{16} \frac{m_t^3\sqrt{2}G_F}
{8\pi^2}\frac{m_t}{-1+\beta^2z^2} \nonumber \\ & &
\Big [2\hat s(a^2\beta^4+\tilde{a}^2)[{\rm Re}C_0(\hat s,
m_t^2,m_t^2,m_t^2)(\hat s-m_{\varphi}^2)
\nonumber \\ & & +{\rm Im}C_0(\hat s,m_t^2,m_t^2,m_t^2)
\Gamma_{\varphi}m_{\varphi}]-4a^2\beta^2(\hat s-m_{\varphi}^2)
\Big ] \nonumber \\& & +\frac{3}{32}\left (\frac{m_t^3\sqrt{2}G_F}{
8\pi^2}\right )^2 \Big [ \hat s^3
(a^2\tilde{a}^2\beta^2+\tilde{a}^4)
\vert C_0(\hat s,m_t^2,m_t^2,m_t^2)\vert^2\nonumber \\ & & +
\hat s(a^4\beta^2+a^2\tilde{a}^2)\vert 2-\hat s\beta^2
C_0(\hat s,m_t^2,m_t^2,m_t^2)\vert^2 \Big ]
\Big \},\label{11}\end{eqnarray}

\begin{eqnarray}c_{g0}&=&c_{g0,{\rm Born}}+c_{g0,{\rm resonance}} \nonumber \\
 c_{g0,\rm Born}&=& \frac{-g_s^4(7 + 9 \beta^2 z^2)}
  {192 E_1^4(-1+\beta^2z^2)^2}
     (E_1^4 -
       2 E_1^2  m_t^2 +
       2  m_t^4 \nonumber \\& &-
       2\beta^2E_1^4z^2 +
       2\beta^2E_1^2m_t^2
        z^2 + \beta^4E_1^4z^4) \nonumber \\
c_{g0,{\rm resonance}}&=&\frac{g_s^4}{
(\hat s-m_{\varphi}^2)^2+
\Gamma_{\varphi}^2m_{\varphi}^2}   \Big \{-\frac{1}{16}
\frac{m_t^3\sqrt{2}G_F}{
8\pi^2}\frac{m_t}{-1+\beta^2z^2}\nonumber \\ & &
\Big [2\hat s(a^2\beta^4-\tilde{a}^2)[{\rm Re}C_0(\hat s,m_t^2,m_t^2,m_t^2)
(\hat s-m_{\varphi}^2)
\nonumber \\& & +{\rm Im}C_0(\hat s,m_t^2,m_t^2,m_t^2)
\Gamma_{\varphi}m_{\varphi}]-4a^2\beta^2(\hat s-m_{\varphi}^2)
\Big ] \nonumber \\& & +\frac{3}{32}\left (\frac{m_t^3\sqrt{2}G_F}{
8\pi^2}\right )^2 \Big [\hat s^3(a^2\tilde{a}^2\beta^2-\tilde{a}^4)
\vert C_0(\hat s,m_t^2,m_t^2,m_t^2)\vert^2 \nonumber \\ & & +
\hat s(a^4\beta^2-a^2\tilde{a}^2)\vert 2-\hat s\beta^2
C_0(\hat s,m_t^2,m_t^2,m_t^2)
\vert^2 \Big ]
\Big \},\label{12}\end{eqnarray}

\begin{eqnarray} c_{g4}=\frac{g_s^4\beta^2\left( 7 + 9{{\beta }^2}
{z^2} \right)(1-z^2)}{32
   ( -1 +\beta^2 z^2)^2}
      ,\end{eqnarray}

\begin{eqnarray}c_{g5}&=&c_{g5,{\rm Born}}+c_{g5,{\rm resonance}}\nonumber \\
 c_{g5,\rm Born}&= &
\frac{ g_s^4\beta ^2 \left( 7 + 9 \beta ^2z^2 \right)}
{ {96 {{{E_1}}^2}
     {{\left( {E_1} + {m_t} \right) }^2}
      {{\left( -1 + {{\beta }^2}{z^2} \right) }^2}}}
         ( {{{E_1}}^4} + 2 {{{E_1}}^3} {m_t} -
       {{{E_1}}^2} {{{m_t}}^2}  \nonumber \\& & -
       4 {E_1} {{{m_t}}^3} - 2 {{{m_t}}^4}-
       2 {{\beta }^2} {{{E_1}}^3} {m_t} {z^2}-
       2 {{\beta }^2} {{{E_1}}^2} {{{m_t}}^2}
        {z^2} - {{\beta }^4} {{{E_1}}^4} {z^4}  )\nonumber \\
 c_{g5,{\rm resonance}}&=&\frac{g_s^4}{
(\hat s-m_{\varphi}^2)^2+
\Gamma_{\varphi}^2m_{\varphi}^2}\Big \{\frac{1}{16}\frac {m_t^3\sqrt{2}G_F}{
8\pi^2}\frac{m_t}{-1+\beta^2z^2} \nonumber \\ & &
\Big [4\hat s a^2\beta^4[{\rm Re}C_0(\hat s,m_t^2,m_t^2,m_t^2)
(\hat s-m_{\varphi}^2)
\nonumber \\& & +{\rm Im}C_0(\hat s,m_t^2,m_t^2,m_t^2)
\Gamma_{\varphi}m_{\varphi}]-8a^2\beta^2(\hat s-m_{\varphi}^2)
\Big ] \nonumber \\& & -\frac{3}{32}\left (\frac{m_t^3\sqrt{2}G_F}{
8\pi^2}\right )^2 \Big [ 2\hat s^3 a^2\tilde{a}^2\beta^2
\vert C_0(\hat s,m_t^2,m_t^2,m_t^2)\vert^2 \nonumber \\& & +
2\hat s a^4\beta^2 \vert 2-\hat s\beta^2
C_0(\hat s,m_t^2,m_t^2,m_t^2)\vert^2 \Big ]
\Big \},\label{14}\end{eqnarray}

\begin{eqnarray} c_{g6}=
{\frac{g_s^4{{\beta }^4} z \left( 7 + 9 {{\beta }^2} {z^2}
                           \right)(z^2-1)}{
   {96\left( {E_1} + {m_t} \right)
     {{\left( -1 + {{\beta }^2} {z^2} \right) }^2}}}}.\end{eqnarray}

\noindent The scalar three point function $C_0$ appearing in equations
(\ref{11}),
(\ref{12}) and (\ref{14}) is given by

\begin{eqnarray}& & C_0(\hat{s},m_t^2,m_t^2,m_t^2)=\nonumber \\ & &
\int \frac{d^4 l}{ i\pi^2}
{1\over l^2-m_t^2+i\epsilon}{1\over (l-p_1)^2-m_t^2+i\epsilon}
{1\over (l-p_1-p_2)^2-m_t^2+i\epsilon}.\label{C02}\end{eqnarray}

\noindent Numerically, $R^g_{{\rm even}}$ is dominated by the Born
contributions. This completes our discussion of $R^g_{{\rm even}}$.
\par The CP--violating part $R_{CP}^g$ results from the interference of
the Born diagrams with the amplitudes of figs. 1c--1h (with couplings
$a\tilde{a}=-\gamma_{CP}$) and the interference of the CP--even and
--odd amplitudes of fig. 1h. We found that if $m_{\varphi}$ is of the
 order
of $2m_t$ or larger, $R_{CP}^g$ is dominated in the resonance region by
the contributions from fig. 1h. Since the complete expressions are
rather lengthy, we have split them with respect to the contributions
from the individual diagrams. For example, $b_{g2}^{({\rm h})}$ means
     the
contribution from fig. 1h to the function $b_{g1}^{CP}$ and
$c_{g1}^{({\rm d,c})}$ denotes the part of $c_{g1}$ that is generated by
the diagrams of figs. 1d and 1e.
\par The function $b_{g1}^{CP}$ gets nonzero contributions only from
the box diagrams of fig. 1c:

  \begin{eqnarray} b_{g1}^{CP}&=&b_{g1}^{({\rm c})}=\frac{m_t^3\sqrt{2}G_F
\gamma_{CP} }{ 8\pi^2}
\frac{- g_s^4 E_1}{96 (-1+\beta^2 z^2)}\nonumber \\ & &
\Big \{(7+9\beta z) \Big [{\rm Im}D_{s}(\hat{t})(1-\beta^2)
-2{\rm Im}D_{11}(\hat{t})
\beta z E_1^2(1-\beta^2)\nonumber \\ & &
+2({\rm Im}D_{11}(\hat{t})+{\rm Im}D_{21}(\hat{t}))
\beta^3 zE_1^2(z^2-1)
 \Big] \nonumber \\ & &
+(7-9\beta z) \Big [
-{\rm Im}D_{s}(\hat{u})(1-\beta^2)-2{\rm Im}D_{11}(\hat{u})
\beta z E_1^2(1-\beta^2)\nonumber \\ & &
+2({\rm Im}D_{11}(\hat{u})+{\rm Im}D_{21}(\hat{u}))
\beta^3 zE_1^2(z^2-1) \Big] \Big \}.\end{eqnarray}

\noindent In this expression,

 \begin{eqnarray}& & D_s(\hat t)=  D_0(\hat t)(m_t^2-\hat t)
+C_0(\hat s,m_{\varphi}^2,
m_t^2,m_t^2) \nonumber \\ & &
 D_s(\hat u)= D_0(\hat u)(m_t^2-\hat u)+C_0(\hat s,m_{\varphi}^2,
m_t^2,m_t^2),\end{eqnarray}
\noindent (where $C_0(\hat{s},m_{\varphi},m_t,m_t)$ is defined in
(\ref{C01})), $\hat{t}=(p_1-k_1)^2,\ \hat{u}=(p_2-k_1)^2$ and

\begin{eqnarray}& &D_0(\hat{t});D_{\mu}(\hat{t});D_{\mu\nu}(\hat{t})=
\int \frac{d^4 l}{i\pi^2}
\frac{1;l_{\mu};l_{\mu}l_{\nu}}{ l^2-m_{\varphi}^2+i\epsilon}\frac{1}
 {(l+k_1)^2-m_t^2+i\epsilon}
\nonumber \\& &\hskip 2cm \frac{1}{(l+k_1-p_1)^2-m_t^2+i\epsilon}
\frac{1}{(l+k_1-p_1-p_2)^2-m_t^2+i\epsilon}\nonumber \\ & & \ \nonumber \\
& & D_{\mu}(\hat{t})= D_{11}(\hat{t})k_{1\mu}-D_{12}(\hat{t})p_{1\mu}
-D_{13}(\hat{t})p_{2\mu} \nonumber \\ & & \ \nonumber \\
& & D_{\mu\nu}(\hat{t})=D_{21}(\hat{t})k_{1\mu}k_{1\nu}
+D_{22}(\hat{t})p_{1\mu}p_{1\nu}
+D_{23}(\hat{t})p_{2\mu}p_{2\nu}\nonumber \\& &\hskip 1cm
-D_{24}(\hat{t})k_{1\mu}p_{1\nu}
-D_{25}(\hat{t})k_{1\mu}p_{2\nu}
+D_{26}(\hat{t})p_{1\mu}p_{2\nu}+D_{27}(\hat{t})g_{\mu\nu}.
\label{DMN}\end{eqnarray}

\noindent $D_0(\hat{u});D_{\mu}(\hat{u});D_{\mu\nu}(\hat{u})$ are obtained
from (\ref{DMN}) by interchanging $p_1$ and $p_2$. The functions
$D_{11},\ldots,D_{27}$ can be reduced to expressions which contain only
the scalar two--, three-- and four--point functions $B_0,\ C_0,\ D_0$
(see e.g.\cite{PV} ).\par The function $b_{g2}^{CP}$ reads:

\begin{eqnarray}b_{g2}^{CP}&=&b_{g2}^{\rm (c)}+b_{g2}^{\rm (g)}+b_{g2}^{\rm
(h)}\nonumber \\
 b_{g2}^{\rm (c)}&=&\frac{m_t^3\sqrt{2}G_F
\gamma_{CP}}{8\pi^2}\frac{g_s^4 \beta}{96(-1+\beta^2 z^2)} \nonumber \\
& &
\Big \{(7+9\beta z) \Big [
-{\rm Im}D_{s}(\hat{t})\frac{m_t}{E_1+m_t}( E_1+m_t+\beta z E_1)\nonumber
 \\ & &
+2{\rm Im}D_{11}(\hat{t})m_t E_1(-E_1+E_1 z^2-m_t z^2)
+4{\rm Im} D_{27}(\hat t)m_t\nonumber \\ & &
+2({\rm Im}(D_{11}\hat{t})+{\rm Im}D_{21}(\hat{t}))
\beta^2 E_1^2(z^2-1)(2m_t+E_1 z^2-m_t z^2) \Big]\nonumber \\& & +
(7-9\beta z) \Big [
-{\rm Im}D_{s}(\hat{u})\frac{m_t}{E_1+m_t}( E_1+m_t-\beta z
      E_1)\nonumber
\\& &
+2{\rm Im}D_{11}(\hat{u})m_t E_1(-E_1+E_1 z^2-m_t z^2)
+4{\rm Im} D_{27}(\hat u)m_t \nonumber \\& &
+2({\rm Im}D_{11}(\hat{u})+{\rm Im}D_{21}(\hat{u}))
\beta^2 E_1^2(z^2-1)(2m_t+E_1 z^2-m_t z^2)\Big]
\Big \} \nonumber \\
b_{g2}^{\rm (g)}&=&\frac{m_t^3\sqrt{2}G_F
\gamma_{CP}}{ 8\pi^2}
\frac{3 g_s^4 m_t\beta^3 z^2 {\rm Im}G(\hat s)}{8(-1+\beta^2 z^2)}\nonumber \\
& &\ \ \nonumber \\ & &\ \  \nonumber \\
 b_{g2}^{\rm (h)}&=&\frac{1}{
(\hat s-m_{\varphi}^2)^2+
\Gamma_{\varphi}^2m_{\varphi}^2}\frac{m_t^3\sqrt{2}G_F
\gamma_{CP}}{8\pi^2}
\frac{g_s^4 m_t\beta}{4(-1+\beta^2 z^2)}\nonumber \\ & &
\Big \{2m_t^2\Big [ {\rm Re}C_0(\hat s,m_t^2,m_t^2,m_t^2)
\Gamma_{\varphi}m_{\varphi}
\nonumber \\& &-
{\rm Im}C_0(\hat s,m_t^2,m_t^2,m_t^2)
(\hat{s}-m_{\varphi}^2)\Big ]+\Gamma_{\varphi} m_{\varphi}\Big \}.
\end{eqnarray}

\noindent  As can be seen seen explicitly from these formulae,
                                          all contributions to
the functions $b_{g1}^{CP}$ and $b_{g2}^{CP}$ result either from
absorptive parts of the one loop amplitudes or from terms of the form:
width $\Gamma_{\varphi}$ times dispersive terms (which is
                                                 present
                                                 only in $b_{g2}^{({\rm
h})}$).
This is in agreement with the general statements made in section 2.
One can also check the relations following from Bose
symmetry as given in table 1. The functions $c_{g1}$ and $c_{g2}$ arise
from dispersive parts in the one loop amplitude or width terms times
absorptive parts (which is present
                  only in $c_{g2}^{({\rm h})}$ below). They read:

\begin{eqnarray}c_{g1}&=&c_{g1}^{\rm (c)}+c_{g1}^{\rm
                                            (d,e)}+c_{g1}^{\rm (f)}
\nonumber \\
c_{g1}^{\rm (c)}&=&\frac{m_t^3\sqrt{2}G_F
\gamma_{CP}}{8\pi^2}
\frac{-g_s^4 E_1}{96(-1+\beta^2 z^2)}\nonumber \\ & &
\Big \{(7+9\beta z) \Big [{\rm Re}D_{s}(\hat{t})(1-\beta^2)
-2{\rm Re}D_{11}(\hat{t})
\beta z E_1^2(1-\beta^2)\nonumber \\ & &
-2({\rm Re}D_{11}(\hat{t})+{\rm Re}D_{21}(\hat{t}))
\beta^3 zE_1^2(z^2-1)
 \Big] \nonumber \\& &
+(7-9\beta z) \Big [
-{\rm Re}D_{s}(\hat{u})(1-\beta^2)-2{\rm Re}D_{11}(\hat{u})
\beta z E_1^2(1-\beta^2)\nonumber \\& &
-2({\rm Re}D_{11}(\hat{u})+{\rm Re}D_{21}(\hat{u}))
\beta^3 zE_1^2(z^2-1) \Big] \Big \}\nonumber \\
c_{g1}^{\rm (d,e)}&=&\frac{m_t^3\sqrt{2}G_F
\gamma_{CP}}{8\pi^2}\frac{-g_s^4 E_1}{96(-1+\beta^2 z^2)} \nonumber \\&
                                            &
\Big \{(9\beta z+7) \Big [
2C_{0}(\hat{t},m_{\varphi}^2,m_t^2,m_t^2)(\beta^2-1)\nonumber \\& &
+{\beta z\over\beta z-1}C_s(\hat{t})(\beta^2 z^2-2\beta^2+1)
                                    \Big]\nonumber
\\& &
+(9\beta z-7) \Big [
2C_{0}(\hat{u},m_{\varphi}^2,m_t^2,m_t^2)(\beta^2-1)\nonumber \\& &
+{\beta z\over\beta z+1}C_s(\hat{u})(\beta^2 z^2-2\beta^2+1) \Big]
\Big \} \nonumber \\
 c_{g1}^{\rm (f)}&=&\frac{m_t^3\sqrt{2}G_F
\gamma_{CP}}{ 8\pi^2}\frac{-g_s^4}{192E_1(-1+\beta^2 z^2)} \nonumber
            \\& &
\Big \{(7+9\beta z)(B_0(m_t^2,m_{\varphi}^2,m_t^2)
-B_0(\hat t,m_{\varphi}^2,m_t^2))\frac{\beta^2-1}{\beta z-1} \nonumber
\\& &
+(7-9\beta z)(B_0(m_t^2,m_{\varphi}^2,m_t^2)
-B_0(\hat u,m_{\varphi}^2,m_t^2))\frac{\beta^2-1}{\beta z+1}\Big \}
 , \end{eqnarray}
\noindent where we used the notation

\begin{eqnarray}& & C_s(\hat t)=C_0(\hat t,m_{\varphi}^2,m_t^2,m_t^2)+
\frac{B_0(m_t^2,
m_{\varphi}^2,m_t^2)-B_0(\hat t,m_{\varphi}^2,m_t^2)
}{m_t^2-\hat t}\nonumber \\
& & C_s(\hat u)=C_0(\hat u,m_{\varphi}^2,m_t^2,m_t^2)+\frac{B_0(m_t^2,
m_{\varphi}^2,m_t^2)-B_0(\hat u,m_{\varphi}^2,m_t^2)}{
  m_t^2-\hat u },\end{eqnarray}

\noindent and

\begin{eqnarray}& &C_0(\hat{t},m_{\varphi}^2,m_t^2,m_t^2)=
\nonumber \\& &
\int \frac{d^4 l}{i\pi^2}
\frac{1}{l^2-m_{\varphi}^2+i\epsilon}\frac{1}{(l+k_1)^2-m_t^2+i\epsilon}
\frac{1}{(l+k_1-p_1)^2-m_t^2+i\epsilon}
\label{C03}.\end{eqnarray}

\noindent $C_0(\hat{u},m_{\varphi}^2,m_t^2,m_t^2)$ is obtained from (\ref{C03})
by the replacement $p_1\to p_2$.

\noindent Finally,

\begin{eqnarray}c_{g2}&=&c_{g2}^{\rm (c)}+c_{g2}^{\rm (d,e)}+c_{g2}^{\rm
                (f)}
+c_{g2}^{\rm (g)}+c_{g2}^{\rm (h)},\nonumber \\
 c_{g2}^{\rm (c)}&=&\frac{m_t^3\sqrt{2}G_F
\gamma_{CP}}{ 8\pi^2}\frac{-g_s^4 \beta}{96(-1+\beta^2 z^2)} \nonumber
\\& &
\Big \{(7+9\beta z) \Big [
-{\rm Re}D_{s}(\hat{t})\frac{m_t}{E_1+m_t}( E_1+m_t-\beta z
                       E_1)\nonumber
\\& &
+2{\rm Re}D_{11}(\hat{t})m_t E_1(E_1-E_1 z^2+m_t z^2)
+4{\rm Re} D_{27}(\hat t)m_t\nonumber \\& &
+2({\rm Re}D_{11}(\hat{t})+{\rm Re}D_{21}(\hat{t}))
\beta^2 E_1^2(z^2-1)(2m_t+E_1 z^2-m_t z^2) \Big]\nonumber \\& & +
(7-9\beta z) \Big [
-{\rm Re}D_{s}(\hat{u})\frac{m_t}{E_1+m_t}( E_1+m_t+\beta z
                       E_1)\nonumber
\\& &
+2{\rm Re}D_{11}(\hat{u})m_t E_1(E_1-E_1 z^2+m_t z^2)
+4 {\rm Re}D_{27}(\hat u)m_t \nonumber \\ & &
+2({\rm Re}D_{11}(\hat{u})+{\rm Re}D_{21}(\hat{u}))
\beta^2 E_1^2(z^2-1)(2m_t+E_1 z^2-m_t z^2)\Big]
\Big \}\nonumber \\
 c_{g2}^{\rm (d,e)}&=&\frac{m_t^3\sqrt{2}G_F
\gamma_{CP}}{8\pi^2}\frac{-g_s^4 \beta}{96(-1+\beta^2 z^2)} \nonumber
\\& &
\Big \{(7+9\beta z ) \Big [
2C_{0}(\hat{t},m_{\varphi}^2,m_t^2,m_t^2)\frac{m_t}{E_1+m_t}(E_1+m_t-
\beta z E_1)\nonumber \\& &+C_s(\hat t)
\frac{-m_t-E_1 z^2+m_t z^2}{E_1^2(1-\beta z)}
(E_1^2-2 m_t^2-\beta^2 E_1^2 z^2)\Big ]\nonumber \\& &+(7-9\beta z
                                       )\Big [
2C_{0}(\hat{u},m_{\varphi}^2,m_t^2,m_t^2)\frac{m_t}{E_1+m_t}(E_1+m_t+
\beta z E_1)\nonumber \\& &+C_s(\hat u)\frac{-m_t-E_1 z^2+m_t
                                       z^2}{E_1^2(1+\beta
z )}
(E_1^2-2 m_t^2-\beta^2 E_1^2 z^2)\Big ]\Big \}\nonumber \\ & & \
\nonumber \\ & & \ \nonumber \\
c_{g2}^{\rm (f)}&=&\frac{m_t^3\sqrt{2}G_F
\gamma_{CP}}{8\pi^2}\frac{g_s^4}{192(-1+\beta^2 z^2)} \nonumber \\& &
\Big \{(7+9\beta z)(B_0(m_t^2,m_{\varphi}^2,m_t^2)
-B_0(\hat t,m_{\varphi}^2,m_t^2)) \nonumber \\& &\frac{m_t}{E_1+m_t}
\frac{E_1+m_t-\beta z E_1}{E_1^2(1-\beta z)}\nonumber \\& &
+(7-9\beta z)(B_0(m_t^2,m_{\varphi}^2,m_t^2)
-B_0(\hat u,m_{\varphi}^2,m_t^2))\nonumber \\& &\frac{m_t}{E_1+m_t}
\frac{E_1+m_t+\beta z E_1}{E_1^2(1+\beta z)}\Big \} \nonumber \\
c_{g2}^{\rm (g)}&=&\frac{m_t^3\sqrt{2}G_F
\gamma_{CP}}{8\pi^2}
\frac{-3 g_s^4 m_t\beta^3z^2{\rm Re}G(\hat s)}{8( -1+\beta^2 z^2)
               }\nonumber \\
 c_{g2}^{\rm (h)}&=&\frac{g_s^4\gamma_{CP}}{
(\hat s-m_{\varphi}^2)^2+
\Gamma_{\varphi}^2m_{\varphi}^2}\Big \{-\frac{1}{16}
                                     \frac{m_t^3\sqrt{2}G_F
 }{8\pi^2}\frac
{ m_t\beta}{-1+\beta^2 z^2}\nonumber \\& &
\Big [2\hat{s}(1+\beta^2)
 [ {\rm Re}C_0(\hat s,m_t^2,m_t^2,m_t^2)(\hat{s}-m_{\varphi}^2)
\nonumber \\& &+
{\rm Im}C_0(\hat s,m_t^2,m_t^2,m_t^2)\Gamma_{\varphi}m_{\varphi}
]-4(\hat{s}-m_{\varphi}^2)\Big ]\nonumber \\& & +\frac{3}{32}
\left (\frac{m_t^3\sqrt{2}G_F
 }{8\pi^2}\right )^2[2\hat{s}^3\beta\tilde{a}^2\vert C_0(\hat
 s,m_t^2,m_t^2,m_t^2)\vert ^2\nonumber \\& &+
 2\hat s\beta a^2\vert 2-\hat s \beta^2
  C_0(\hat s,m_t^2,m_t^2,m_t^2)\vert ^2 ] \Big \}.
 \end{eqnarray}

\vfill
\eject

\newpage
\noindent {\Large{\bf Table Caption}}
\vskip 0.5cm
\noindent {\bf Table 1}: Transformation properties of the structure
functions defined in (2.2)--(2.3).
\newpage
\renewcommand{\arraystretch}{1.2}
\begin{center} {\large {\bf Table 1}}\vskip 1cm
$$
\begin{array}{|r|r|r|r|r|r|}\hline
\   & \hfil {\rm CP}\hfil   & \hfil {\rm P} \hfil  &\hfil {\rm  T
}\hfil
&\hfil {\rm  CPT}\hfil  &\hfil {\rm ``Bose" }\hfil \\
  \   &  \  &   \
  &\hfil { ({\rm Im}{\cal T}=0)} \hfil
&\hfil{({\rm Im}{\cal T}=0)} \hfil  &\ \\ \hline \hline
A^g(z) & A^g(z) & A^g(z) & A^g(z) & A^g(z) & A^g(-z) \\
b_{g1}^{\pm}(z) & b_{g1}^{\mp}(z) &  -b_{g1}^{\pm}(z) & b_{g1}^{\pm}(z) &
b_{g1}^{\mp}(z) &-b_{g1}^{\pm}(-z) \\
b_{g2}^{\pm}(z) & b_{g2}^{\mp}(z) & -b_{g2}^{\pm}(z) & b_{g2}^{\pm}(z) &
b_{g2}^{\mp}(z) &b_{g2}^{\pm}(-z) \\
b_{g3}^{\pm}(z) & b_{g3}^{\mp}(z) & b_{g3}^{\pm}(z) & -b_{g3}^{\pm}(z) &
-b_{g3}^{\mp}(z) &-b_{g3}^{\pm}(-z) \\
c_{g0}(z) & c_{g0}(z) & c_{g0}(z) & c_{g0}(z) &c_{g0}(z) & c_{g0}(-z)  \\
c_{g1}(z) & -c_{g1}(z) & -c_{g1}(z) & -c_{g1}(z) &c_{g1}(z) & -c_{g1}(-z)  \\
c_{g2}(z) & -c_{g2}(z) & -c_{g2}(z) & -c_{g2}(z) &c_{g2}(z) & c_{g2}(-z)  \\
c_{g3}(z) & -c_{g3}(z) & c_{g3}(z) & c_{g3}(z) &-c_{g3}(z) & -c_{g3}(-z)  \\
c_{g4}(z) & c_{g4}(z) & c_{g4}(z) & c_{g4}(z) &c_{g4}(z) &c_{g4}(-z)  \\
c_{g5}(z) & c_{g5}(z) & c_{g5}(z) & c_{g5}(z) &c_{g5}(z) &c_{g5}(-z)  \\
c_{g6}(z) & c_{g6}(z) & c_{g6}(z) & c_{g6}(z) &c_{g6}(z) &-c_{g6}(-z)  \\
c_{g7}(z) & c_{g7}(z) & -c_{g7}(z) & -c_{g7}(z) &-c_{g7}(z) &c_{g7}(-z)  \\
c_{g8}(z) & c_{g8}(z) & -c_{g8}(z) & -c_{g8}(z) &-c_{g8}(z)
&-c_{g8}(-z)\\ \hline \end{array}$$
\end{center}

\newpage
\noindent {\Large{\bf Figure Captions}}
\vskip 0.5cm
\noindent {\bf Fig. 1}: Born level QCD and $\varphi$ exchange Feynman
diagrams which contribute to the production density matrix for $gg\to
t\bar{t}$. Diagrams with crossed gluons are not shown.
\vskip 0.3cm
\noindent {\bf Fig. 2}: Born level QCD and relevant $\varphi$ exchange
Feynman diagrams for $q\bar{q}\to t\bar{t}$.
\vskip 0.3cm
\noindent {\bf Fig. 3}: Expectation value $\langle \hat{{\bf k}}\cdot  ({\bf
s}_+-{\bf s}_-) \rangle_g$ as a function of the parton CM energy for
$m_{\varphi}=100$ GeV (dashed curve) and $m_{\varphi}=350$ GeV (solid
curve).
\vskip 0.3cm
\noindent {\bf Fig. 4}: Expectation value  $\langle \hat{{\bf k}}\cdot  ({\bf
s}_+\times {\bf s}_-) \rangle_g$ for the same choice of Higgs masses as
in Fig. 3.
\vskip 0.3cm
\noindent {\bf Fig. 5}: Same as Fig. 3, but for $\langle \hat{{\bf k}}\cdot
({\bf
s}_+-{\bf s}_-) \rangle_q$.
\vskip 0.3cm
\noindent {\bf Fig. 6}: Same as Fig. 4, but for $\langle \hat{{\bf k}}\cdot
({\bf
s}_+\times {\bf s}_-) \rangle_q$.
\vskip 0.3cm
\noindent {\bf Fig. 7}: Signal--to--noise ratio for the observable $A_1$
(defined in (4.1)) as a function of the Higgs boson mass $m_{\varphi}$
for proton--proton CM energies $\sqrt{s}=15$ TeV (solid curve) and
$\sqrt s=40$ TeV (dashed curve). Here $m_t=150$ GeV, $a=-\tilde{a}=1,\
\hat{g}_{VV}=1$.
\vskip 0.3cm
\noindent {\bf Fig. 8}: Same as fig. 7, but for the observable $A_2$
(defined in (4.2)).
\vskip 0.3cm
\noindent {\bf Fig. 9}: Same as fig. 7, but for the observable $T_2$
(defined in (4.3)).

 \end{document}